\documentclass[pre,twocolumn,preprintnumbers,amsmath,amssymb]{revtex4-1}

\usepackage{graphicx}
\usepackage{bm}
\usepackage{amssymb, dsfont}
\usepackage{color}

\newcommand{\R}{{\mathbf r}}
\newcommand{\RR}{{\mathbf R}}
\newcommand{\rr}{{\mathbf r}}

\newcommand{\F}{{\mathbf F}}

\newcommand{\Q}{{\mathbf Q}}

\newcommand{\e}{{\hat{\mathbf e}}}

\newcommand{\n}{{\hat{\mathbf n}}}

\newcommand{\N}{\mbox{\boldmath$\nabla$}}
\newcommand{\BEQ}{\begin{equation}}
\newcommand{\EEQ}{\end{equation}}
\newcommand{\nn}{\nonumber}

\newcommand{\beq}{\begin{equation}}
\newcommand{\eeq}{\end{equation}}
\newcommand{\bea}{\begin{eqnarray}}
\newcommand{\eea}{\end{eqnarray}}

\definecolor{darkgreen}{rgb}{0,0.5,0}

\begin{document}

 \begin{@twocolumnfalse}
 \begin{center}
    \mbox{
{
\color{blue} \href{http://www.nature.com/articles/srep34146}{DOI:10.1038/srep34146}
}
    }
\end{center}
  \end{@twocolumnfalse}

%
\title{Shape and Displacement Fluctuations in Soft Vesicles Filled by Active Particles}

\author{Matteo Paoluzzi$^{1,2}$}
\email{mpaoluzz@syr.edu}
\author{Roberto Di Leonardo$^{2,3}$}
\author{M. Cristina Marchetti$^1$}
\author{Luca Angelani$^{2,4}$}

\affiliation{$^1$ Department of Physics and Syracuse Soft Matter Program, Syracuse University, Syracuse NY 13244, USA \\
$^2$Dipartimento di Fisica Universit\`a Sapienza, P.le A Moro 2, 00185 Rome, Italy \\
$^3$NANOTEC-CNR, Institute of Nanotechnology, Soft and Living Matter Laboratory, Piazzale A. Moro 2, I-00185, Roma, Italy \\
$^4$ISC-CNR, Institute for Complex Systems, Piazzale A. Moro 2, I-00185 Roma, Italy}


\begin{abstract}
We investigate numerically the dynamics of shape and displacement
fluctuations of two-dimensional flexible vesicles  filled with active particles.
At low concentration most of the active particles accumulate at the
boundary of the vesicle where positive  particle number fluctuations are amplified
by trapping, leading to the formation of  pinched spots of high density, curvature and pressure. 
At high concentration the active particles cover the vesicle boundary almost uniformly, resulting in  fairly homogeneous
pressure and curvature, and nearly circular vesicle shape.
The change  between polarized and spherical shapes
is driven by the number of active particles.
The center-of-mass of the vesicle 
performs a persistent random walk with a long time diffusivity that is
strongly enhanced for elongated active particles due to
orientational correlations in their direction of propulsive motion.
 In our model shape-shifting induces directional
sensing and the cell spontaneously migrate along the polarization
direction.
\end{abstract}

\flushbottom
\maketitle
\thispagestyle{empty}

\section*{Introduction}
Active systems are collections of 
 agents that convert the energy of the environment in systematic movement  \cite{Vicsek12,Marchetti13,Cates12}.
Examples include bacterial colonies \cite{Peruani12}, epithelial cell layers \cite{Bi15},
self-propelled colloids \cite{Palacci10}, swimming microorganisms \cite{Berg04}, schools of fish \cite{Hemeltijk08} 
and bird flocks \cite{Ballerini08}.
Active particles can form gas, liquid, liquid crystal or glassy phases with structural properties
remarkably similar to those of ordinary materials \cite{Marchetti15,Henkes11,Render13,Berthier14,Berthier13,Tailleur08,Marchetti15,Zhou14}. 
Active systems are, however, out-of-equilibrium. Hence their steady state is not described by the Boltzmann distribution
and they can support spontaneous, self-sustained motion, which can in turn
 be enhanced, stabilized or suppressed by suitably designed confining geometries \cite{Hol14,Sanchez12,wioland,Galajda14}. 
It has been shown that active agents can give rise to ratchet effects, \cite{Galajda07,Wan08,Angelani11,Reichhardt16} power microgears \cite{Angelani09,DiLeonardo10,Sokolov10}, drive spontaneous accumulations of passive colloids over target regions \cite{Koumakis13}, and exhibit 
long lived density fluctuations \cite{Narayan07}.
From a theoretical point of view, the effect of confinement
has been used to investigate  the concept
of pressure in active systems \cite{Solon15,Solon15b,Yang14,Takatori14} 
and the effect of wall curvature on both active particles
 \cite{Smallenburg15} and  passive tracers \cite{Mallory14}.
Strong confinement can induce hysteretic dynamics \cite{Fily15} or sustained spontaneous density oscillations \cite{Paoluzzi15}.
The role of curved walls on active gas has been investigated in Ref.\cite{Fily14}.

Previous work has focused on confinement  by rigid walls.
While recent studies have investigated the effects of active baths on flexible open chains \cite{har2014,kai2014,Shin15},
the case of swimmers confined by deformable boundaries 
has recently been analyzed only for case of spherical active Brownian particles by Tian {\itshape et al.} \cite{Tian15}.
 An interesting
example of active colloidal cell driven by micro rotators has been theoretically investigated in \cite{Spellings15}.
Here we consider an active vesicle in two dimensions composed 
by a flexible one dimensional membrane enclosing active particles representing an active solute.
The corresponding equilibrium system would be a vesicle filled with a suspension and bounded by a flexible membrane that is permeable to the solvent but not to the solute molecules. In this case, the solute concentration would be uniform throughout the vesicle interior and exert a homogeneous pressure on the membrane whose equilibrium configuration would be spherical, or circular in two dimensions. 
When the solute molecules are active particles or microswimmers, we find that only for high densities of active particles the membrane shape fluctuates around a circle. When the swimmers packing fraction falls below a characteristic value, depending on particles shape,  the vesicle acquires an asymmetric shape characterized by  a bimodal distribution of the local curvatures, with a high curvature peak and a near zero curvature component.  This effect is driven by a feedback mechanism coupling swimmers density and membrane curvature through local pressure. A local fluctuation of particle density produces a local pressure increase that induces a larger curvature on the flexible membrane. Since active particles tend to accumulate at concave boundaries, this local curvature increase drives  further accumulation of swimmers, which in turn raises the local pressure.
The presence of this feedback mechanism is confirmed by a strong correlation between  
the local swimmers density  (or local pressure on the membrane) and the local curvature of the membrane.
Finally, we examine the center of mass dynamics of the whole vesicle and show that
it performs a persistent random walk with a long time diffusivity that is larger for elongated swimmers due to orientational correlations. Interesting, the resulting migratory behavior shares some similarities
with Eukaryotic directed cell migration \cite{Swaney10,Devreotes03}.

\section*{Methods}
We perform two dimensional simulations of $N_s$ {\it run-and-tumble} 
swimmers of width $a$ and length $\ell$ (aspect ratio $\alpha=a/ \ell$)
confined by a deformable membrane. 
We specifically consider swimmers of two different aspect ratios, $\alpha=1/2$ (elongated) and $\alpha=1$ 
(spherical).

\begin{figure}[!t]
\centering
\includegraphics[width=.47\textwidth]{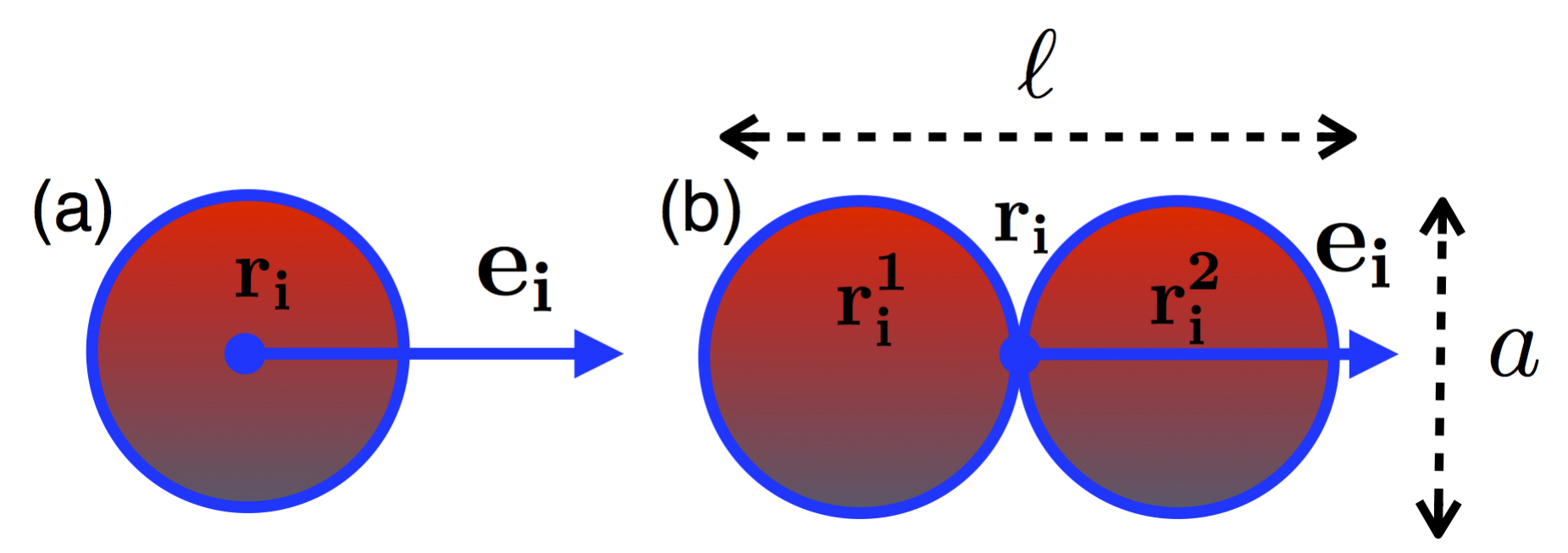}
\caption{ {\bf Pictorial representation of the swimmers.} 
Each swimmer consists of $p$ spherically symmetric force centers aligned along the swimming direction $\mathbf{e}_i$, with $p\!=\!1$ describing  spherical particles
(panel a) and  $p\!=\!2$ elongated ones (panel b).}
\label{fig:mod}       
\end{figure}
%

\subsection*{Swimmers}
We consider $N_s$ run-and-tumble particles in two dimensions. 
The model is the same used in \cite{Angelani09,Angelani11c,Angelani10c,Paoluzzi13,Paoluzzi14}
to describe {\itshape E. coli} bacterial suspensions.
Each particle  consists of a chain of 
$p$ rigidly connected disks  of diameter $\ell/2$ aligned along the swimming direction $\e_i$.
We denote by $\mathbf{r}_i$ the center of mass of the $i$th
swimmer. The position   $\mathbf{r}_i^\beta$, with $\beta=1,...,p$,  of the $\beta$-th disk on the $i$-th swimmer is then
\beq
\mathbf{r}_i^\beta = \mathbf{r}_i + \delta \rr^\beta_i \,   \, .
\eeq
Here we consider $p=1$, corresponding to spherical swimmers with $\delta \rr_i^1=\mathbf{0}$, and $p=2$, corresponding to elongated swimmers, with
$\delta \rr_i^1=-\e_i \ell/4$ and $\delta \rr_i^2=+ \e_i \ell/4$ (panel (b) in Fig. \ref{fig:mod}). 
We assume swimmers interact only through steric repulsion and that the interaction potential is written as the sum of radially symmetric potentials centered at each disk. For this reasons the individual disks that compose our swimmers are also referred to as force centers.
At low Reynolds number, the equations of motion of the $i$-th swimmer  are
\bea\label{eqm}
\mathbf{v}_i &=& \mathbf{M}_i \, \cdot\mathbf{F}_i \;,\\ \nn
\boldsymbol{\omega}_i &=&\mathbf{K}_i  \,\cdot \mathbf{T}_i \;,
\eea
where $\mathbf{v}_i$ is the center of the mass velocity and $\boldsymbol{\omega}_i$ the angular velocity of the $i-$th swimmer. $\mathbf{M}_i$ and $\mathbf{K}_i$ are the translational and rotational mobility matrices 
\bea
\mathbf{M}_i &=& m_{\parallel} \mathbf{\hat{e}}_i  \otimes \mathbf{\hat{e}}_i  + m_{\perp} \left( \mathbf{1} -  \mathbf{\hat{e}}_i  \otimes \mathbf{\hat{e}}_i \right)   \\ \nn
\mathbf{K}_i  &=&  k_{\perp} \left( \mathbf{1} - \mathbf{\hat{e}}_i    \otimes \mathbf{\hat{e}}_i \right) \, ,
\eea
the symbol $\otimes$ is the dyadic product and $\mathbf{1}$ the identity matrix.
In Eq. (\ref{eqm}), $\mathbf{F}_i$ and $\mathbf{T}_i$ are the total force and the
total torque acting on the of the $i$-th swimmer, given by
\bea\label{fandt}
 \mathbf{F}_i  &=& f_0 \mathbf{\hat{e}}_i(1 - \sigma_i) + \sum_{j \neq i, \alpha, \beta} \mathbf{f}(\mathbf{r}_i^\alpha - \mathbf{r}_j^\beta) + \sum_{\alpha} \mathbf{f}_{ext}(\rr_i^\alpha)\\ \nn
 \mathbf{T}_i  &=& \mathbf{t}_r^i \sigma_i +   \sum_{j \neq i, \alpha, \beta} \delta \rr^\alpha_i \times \mathbf{f}(\mathbf{r}_i^\alpha - \mathbf{r}_j^\beta) + \sum_{\alpha} \delta \rr^\alpha_i \times \mathbf{f}_{ext}(\rr_i^\alpha) \, .
\eea
The index $j=1,\dots,N_s$ runs over swimmers, the indices $\alpha=1,..,p$ and $\beta=1,...,p$ run over disks, and
$\sigma_i$ is a state variable, with value $0$ for running swimmers and $1$ for tumbling ones.
During the running state each swimmer
is self-propelled along $\e_i$ with self-propulsion speed $v=m_{\parallel} f_0$. 
In the tumbling state, the random torque $\mathbf{t}_r^i$ rotates
the swimming direction $\e_i$
at the  tumbling rate, $\lambda$.
Moreover, it takes a finite time $(\lambda 10)^{-1}$ for the swimmers 
to reorient the swimming direction.  The external force  $\sum_{\alpha} \mathbf{f}_{ext}(\rr_i^\alpha)$ in Eq. (\ref{fandt})
represent the interaction of the swimmers with the flexible confining boundary. The details of this interaction will be specified in the next section.
Finally, the repulsive force $\mathbf{f}(\mathbf{r})$ is conservative and generated by the potential $V(r) = \frac{a f_0}{12} \ \left( \frac{a}{r} \right)^{12}$\cite{Angelani09}. 
Below we choose units such that  $\ell=m_\parallel=f_0=1$ and use $\lambda=0.1$, 
 $k_{\perp}=4.8$ and $m_{\perp}=0.87$.

%
%
\subsection*{Membrane}

The bounding membrane is modeled as a ring of $N_b$ colloidal beads of diameter $a$ 
connected by springs.
Denoting with $\RR_n$ the position of $n$-th  bead, 
the equation of motion of the membrane in the low Reynolds number regime is given by
\bea\label{motion}
\dot{\RR}_n &=& \mu_b \F_n\;,\hspace{0.2in} \F_n = -\N_n \varphi(\{\RR\}  ,\{\R\} )\;,
\eea
where the potential $\varphi(\{\RR\} ,\{\R\} )$ consists of harmonic and repulsive parts, $\varphi(\{\RR\}  ,\{\R\} ) = \varphi(\{\RR\})^{harm}+\varphi(\{\RR\}  ,\{\R\} )^{rep}$, with
\bea
&&\varphi(\{\RR\})^{harm} = \frac{k}{2}\sum_{n=1}^{N_b}\left( | \RR_{n+1} - \RR_n | - a 
\right)^2 \label{phiarm} \, ,\\ \nn
&&\varphi(\{\RR\}  ,\{\R\} )^{rep} = \sum_{n<m} V(|\RR_n - \RR_m|) + \sum_{n,i,\beta} V(|\RR_n -\R_i^\beta|)\, ,
\eea
where $\RR_{N_b+1}=\RR_1$ in the sum in Eq.~(\ref{phiarm}).
We choose $\mu_b\!=\!\mu$, $k=5\cdot10^2$. 
The external force in Eq. (\ref{fandt}) is  $\mathbf{f}_{ext}(\mathbf{r}_\alpha^i)=-\frac{\partial\varphi(\{\RR\},\{\R\})^{rep}}{\partial\mathbf{r}_\alpha^i}$.

The initial configuration of the membrane is a circle of radius $R_0\!=\!a (2\sin(\pi/N_b))^{-1}$ and area $A_{ref}\!=\!\pi R_0^2$.
The swimmers cover a  fraction $\phi\!=\!N_s a_{swim} / A_{ref}$ of the  initial area of the vescicle, with
$a_{swim}\!=\!p \pi (a/2)^2$ the area of one swimmer. 
The entire vesicle moves in a two dimensional box of side $70\ \ell$ with periodic boundary conditions.
We have performed numerical simulations of membranes composed of $N_b=50, 100, 150$ beads enclosing $N_s$
elongated swimmers ($p=2$) with packing fraction from $\phi=0.07$ up to $0.83$
and $N_s$ spherical swimmer ($p=1$)  with packing fraction from $\phi=0.05$ to $0.82$. Specifically, in the case of elongated swimmers we have simulated systems with $N_s=12,21,32,37,52,69,80$ for $N_b=50$, $N_s=52,69,80,97,112,137,156,208,225,256,316,384,421,448$ for $N_b=100$, and $N_s=80,112,156,208,256,316$, $384,448,540,616,716,812,973$ for $N_b=150$. For spherical swimmers we have used
$N_s=12,21,32,52,80,112,156,208$ for $N_b=50$, $N_s=52,112,208,316,448,616,812$ for $N_b=100$, and
$N_s=316,448,616,812,1020,1264$ for $N_b=150$.

To quantify the shape of the membrane we measure the gyration tensor $\Q$, given by
\BEQ
\Q = \frac{1}{N_b}\sum_{n=1}^{N_b} \left( \RR_n - \RR_{cm}\right)\otimes \left( \RR_{ n}- \RR_{ cm}\right)\;,
\EEQ
with $\RR_{cm}$ the center of the mass of membrane beads.  
From the average values of the two eigenvalues $\lambda_1$ and $\lambda_2$ of $\Q$ we compute the squared radius of gyration $R_g^2=Tr \, \Q$ that gives a measure of the extension of the cell,
\BEQ
R_g^2=\lambda_1+\lambda_2 \;,
\EEQ
and the asphericity~\cite{Aronovitz86} 
\BEQ
\Delta = \frac{\left( \lambda_1 - \lambda_2 \right)^2}{\left( \lambda_1 + \lambda_2 \right)^2}\, .
\EEQ
The value $\Delta=0$ corresponds to a circle and $\Delta=1$ to  a rod.
Since the gyration tensor is a dynamical quantity, the observables
$R_g$ and $\Delta$ are computed from the time average of the eigenvalues.

We characterize the local shape of the membrane
by measuring the local curvature, $\kappa$,
defined as \cite{Sokolov01}
\BEQ
\kappa = \frac{(\RR^{\prime} \times \RR^{\prime\prime})_z}{|\RR^{\prime}|^3} \;,
\EEQ
where $\RR$ is the vector position of a membrane point, 
$\RR^{\prime}$ and $\RR^{\prime\prime}$ are the first and second derivatives of $\RR$ with respect to the  membrane contour length.
Curvature values are evaluated at the beads position along the membrane, using discrete form of the derivatives.
To evaluate the pressure $P$ on the
$n-$th bead, we have computed numerically  the total force that swimmers exert along the local normal 
$\n_n$ to the membrane divided by the average length of the segments connecting such a bead to its neighbors.

\section*{Results}
\begin{figure*}[!t]
\centering
\includegraphics[width=1.\textwidth]{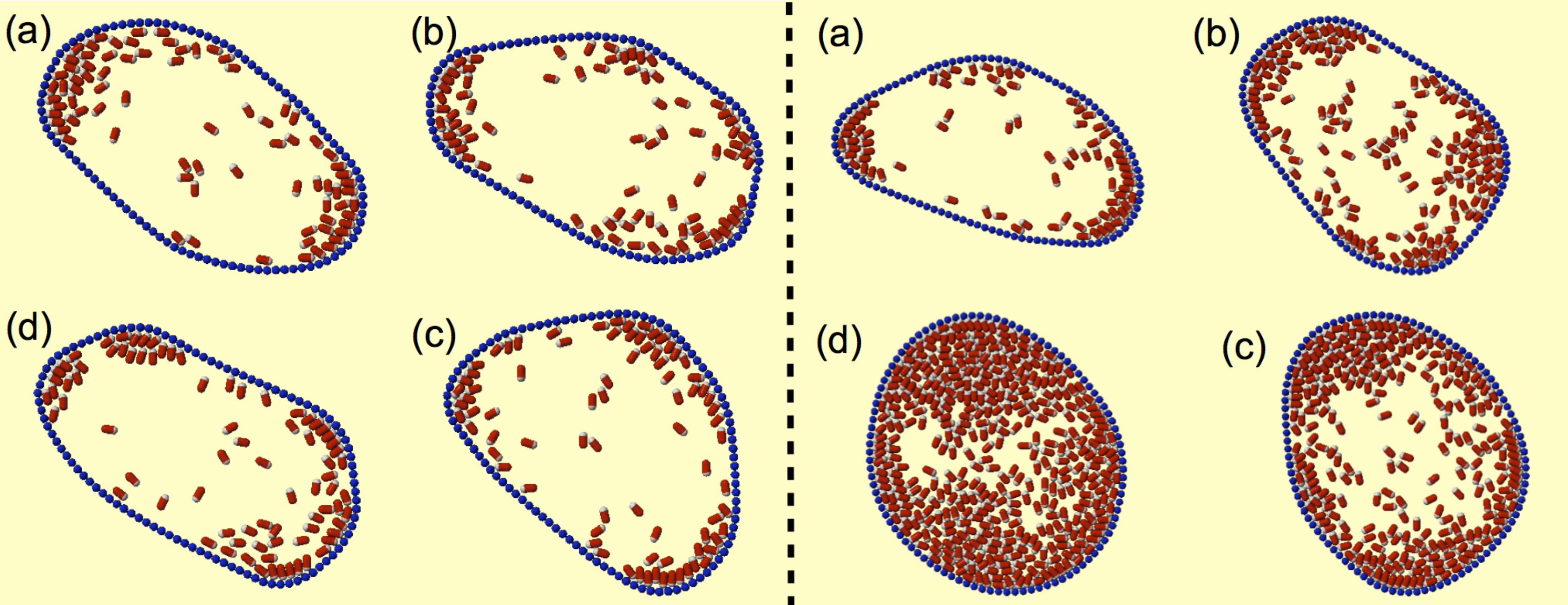}
\caption{ 
{\bf Shape fluctuations.} The bounding membrane is composed of
$N_b=100$ beads. Left panel: snapshots of vesicle shapes explored by the active vesicle for low packing fraction of  elongated swimmers ($\phi=0.16$). Right panel: the vesicle becomes more symmetric as the number  of  active particles increases, 
 $\phi$ is $0.16$ (a), $0.31$ (b), $0.51$ (c) and $0.76$ (d).
}
\label{fig:fig1}       
\end{figure*}

It is well established in the literature that confined active particles 
tend to accumulate along the confining walls~\cite{Tailleur08,Wensink08,Yang14}.
In our case the confining walls are flexible
and swimmer accumulation induces strong distortions of the bounding membrane. 
These distortions are evident in the
snapshots shown in Fig. \ref{fig:fig1} where elongated swimmers are bounded by a membrane of $N_b=100$ beads. 
For low packing fraction (left panel) the membrane
explores different shapes characterized by regions of high curvature.
As the number of swimmers is increased (right panel), the imbalance of particles along the
flexible walls becomes less dramatic and the vesicle assumes more symmetric shapes.

\subsection*{Pressure and global shape properties}
We first discuss  the case of spherical swimmers ($p=1$, aspect ratio $\alpha=1$). 
In this case particle reorientations are solely due to tumbles and no aligning interactions exists between swimmers or swimmers and walls.

\begin{figure*}[!t]
\centering
\includegraphics[width=.95\textwidth]{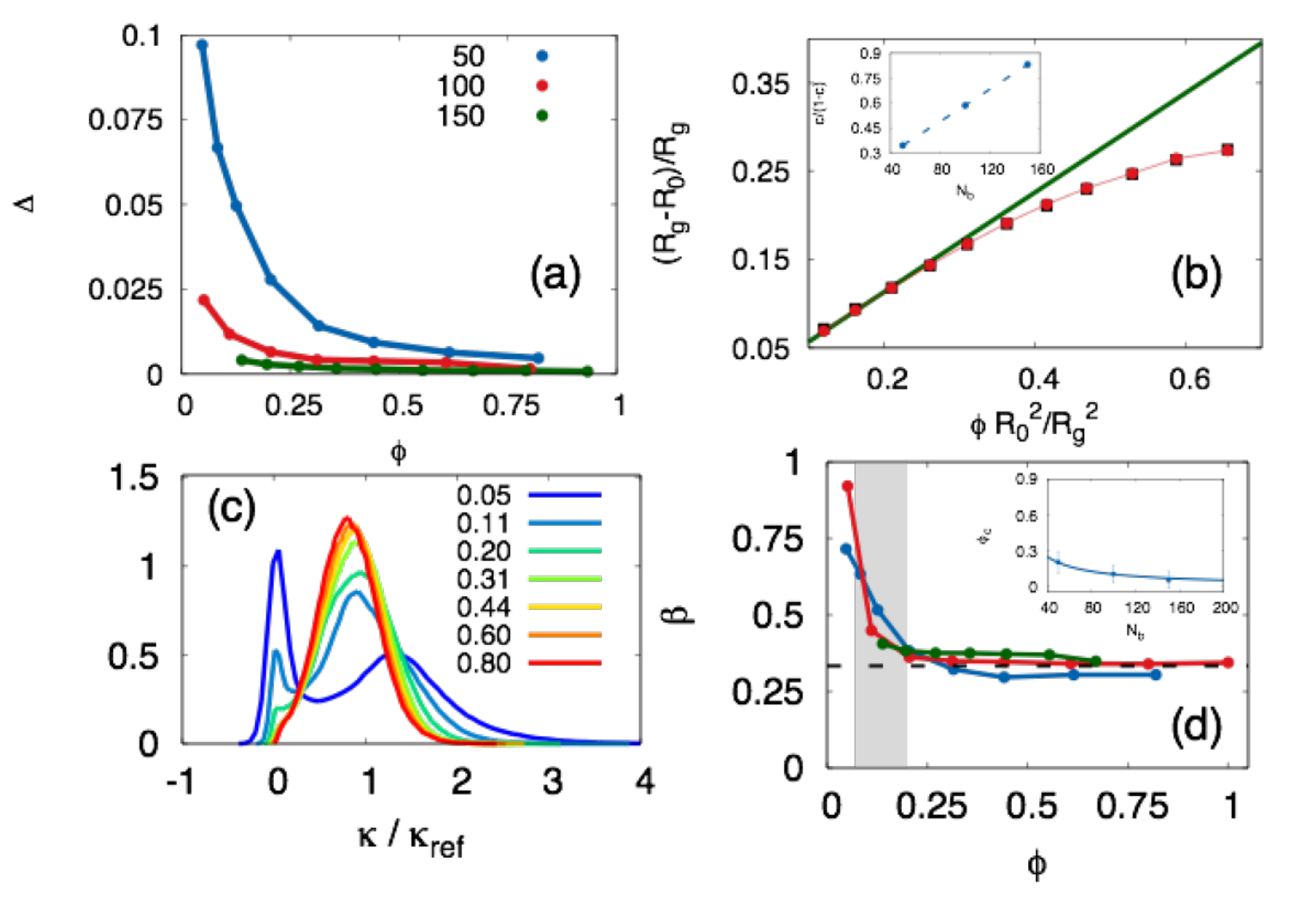}
\caption{{\bf Membrane shape for spherical swimmers.} 
(a) Asphericity parameter for $N_b=50$ (blue symbols), $N_b=100$ (red symbols), and $N_b=150$ (green symbols), the lines are a guide to the eye. The membrane approaches a circular shape with increasing $\phi$.
(b) The red circles are $(R_g-R_0)/R_g$ (the red line is a guide to the eye), the black symbols represent $P N_b / 2 \pi k$, and the green line is the fit to Eq. (\ref{press}). The data are plotted as a function of the area fraction computed with respect the circle of radius $R_g$ for $N_b=150$. Inset: the quantity $c/(1-c)$ as a function of $N_b$.
 (c) Probability distributions of the local curvatures for $N_b=100$ for increasing $\phi$ from $0.05$ (blue) to $0.80$ (red). (d) Parameter $\beta$ as a function of $\phi$ for $N_b=50,100,150$ (blue, red and green), the black dashed line is $\beta$ for a Gaussian distribution. The grey area represents the estimated $\phi_c$ range. Inset: $\phi_c$ obtained from the decoupling between pressure and deformation (blue symbols) the line is the estimate of $\phi_c$ given by $\phi(N_b=N_s)$.}
\label{fig:fig2}       
\end{figure*}

To quantify the deviations of the active vesicle from circular shape we display in Fig.~\ref{fig:fig2}-a the asphericity $\Delta$ for different values of $N_b$ as a function of the swimmers area fraction. 
We find that $\Delta$ rapidly decays to zero with increasing $\phi$ especially for large vesicles ($N_b=150$), indicating
that at high density of swimmers the active vesicle approaches an average  circular shape.
In contrast, we observe deviation from a circular shape for small vesicles in the dilute regime.

Now we quantify the membrane stretching for $N_b=150$ (in this case $\Delta\sim0$ in the whole $\phi$ range explored).
We show in Fig.~\ref{fig:fig2}-(b) that the gyration radius, 
$R_g$ increases with $N_s$. This is true for all vesicle sizes ($N_b=50,100$ not shown in figure), indicating that the active particles exert a pressure that stretches the bounding membrane. 
A simple estimate for the dependence of $R_g$ on swimmer packing fraction can be obtained for a dilute gas of spherical {\itshape run-and-tumble} swimmers. In two dimensions
the pressure of an ideal active gas  of $N_s$ spherical swimmers in an area $A$ is the
so-called swim pressure \cite{Takatori14,Yang14}, given by
\BEQ\label{swim}
P_{swim} =\frac{N_s}{A} \frac{v^2}{2 \lambda \mu}=\frac{\phi}{a_{swim}} \frac{v^2}{2 \lambda \mu} \, ,  
\EEQ
where we have expressed $P_{swim}$ in terms of the initial packing fraction $\phi$.
In presence of confining structures 
the pressure in the bulk is strongly affected
by the finite size effects \cite{Yang14,Solon15b,Marchetti15,Maggi15}.

For example, in the case of one dimensional gas of {\itshape run-and-tumble} particles confined in a box of side $L$ we can write\cite{Maggi15}:
\BEQ\label{conf}
P_{ 1d-box}(L)=  \frac{P_{swim}}{1+2 v/ \lambda L}  \,,
\EEQ

We assume that the internal pressure is responsible of an isotropic deformation of the vesicle from a circle
of radius $R_0$ to a circle of radius $R_g$.
In the dilute regime, we assume that Eq. (\ref{conf}) 
can be recast phenomenologically as
\BEQ\label{pre}
P(R)= \frac{P_{swim}}{1+\mathcal{L}/R} \;,
\EEQ
with $\mathcal{L}$ a fitting parameter. 
A relation between the internal active pressure and the radius $R_g$ in the deformed configuration can be obtained
as follows.
Since the membrane is composed by elastic springs it will store an elastic energy given by,
\BEQ\label{ela}
E(R_g)=\frac{1}{2}\frac{k}{N_b}\left[ 2 \pi (R_g-R_0) \right]^2 \, .
\EEQ
The membrane tension exerts an inward pressure that has to be balanced by the pressure exerted by the active particles, requiring 
\BEQ \label{press}
P(R_g)=\frac{1}{2\pi R_g}\frac{\partial}{\partial R_g} E(R_g)= \frac{2 \pi k}{N_b} \frac{R_g - R_0}{R_g} \, ,
\EEQ
In Fig. \ref{fig:fig2}-(b), the red circles represent the quantity $(R_g-R_0)/R_g$ as a function of the actual area fraction computed as $\phi R_0^2/R_g^2$. 
For $N_b=150$ the membrane has a nearly circular shape so that (\ref{press}) holds and $(R_g-R_0)/R_g$ becomes proportional to the average pressure exerted by the swimmers. This is confirmed by plotting on the same graph  the swimmers pressure as obtained from simulations and divided by $2\pi k/N_b$ (black squares). In the dilute regime, Eq. (\ref{pre})
holds, as a result
the pressure should scale linearly with the packing fraction, provided the correction term $c(R)$ and thus $R$ does not change significantly with the packing fraction. By fitting the low $\phi$ data in Fig. \ref{fig:fig2}-(b) we obtain $c=0.46$. Deviations of the pressure from the linear regime, due to the excluded volume effects, are visible at high $\phi$ \cite{Yang14,Solon15b,Marchetti15}. 
We can repeat the same procedure for membranes of different contour length and obtain $c$ values for different $N_b$.
Assuming $c=(1+\mathcal{L}/R)^{-1}$ we expect that the quantity $c/(1-c)$ should scale linearly with $R$ which is approximately proportional to $N_b$ (see inset of Fig. \ref{fig:fig2}-(b)). 

\begin{figure*}[!t]
\centering
\includegraphics[width=.85\textwidth]{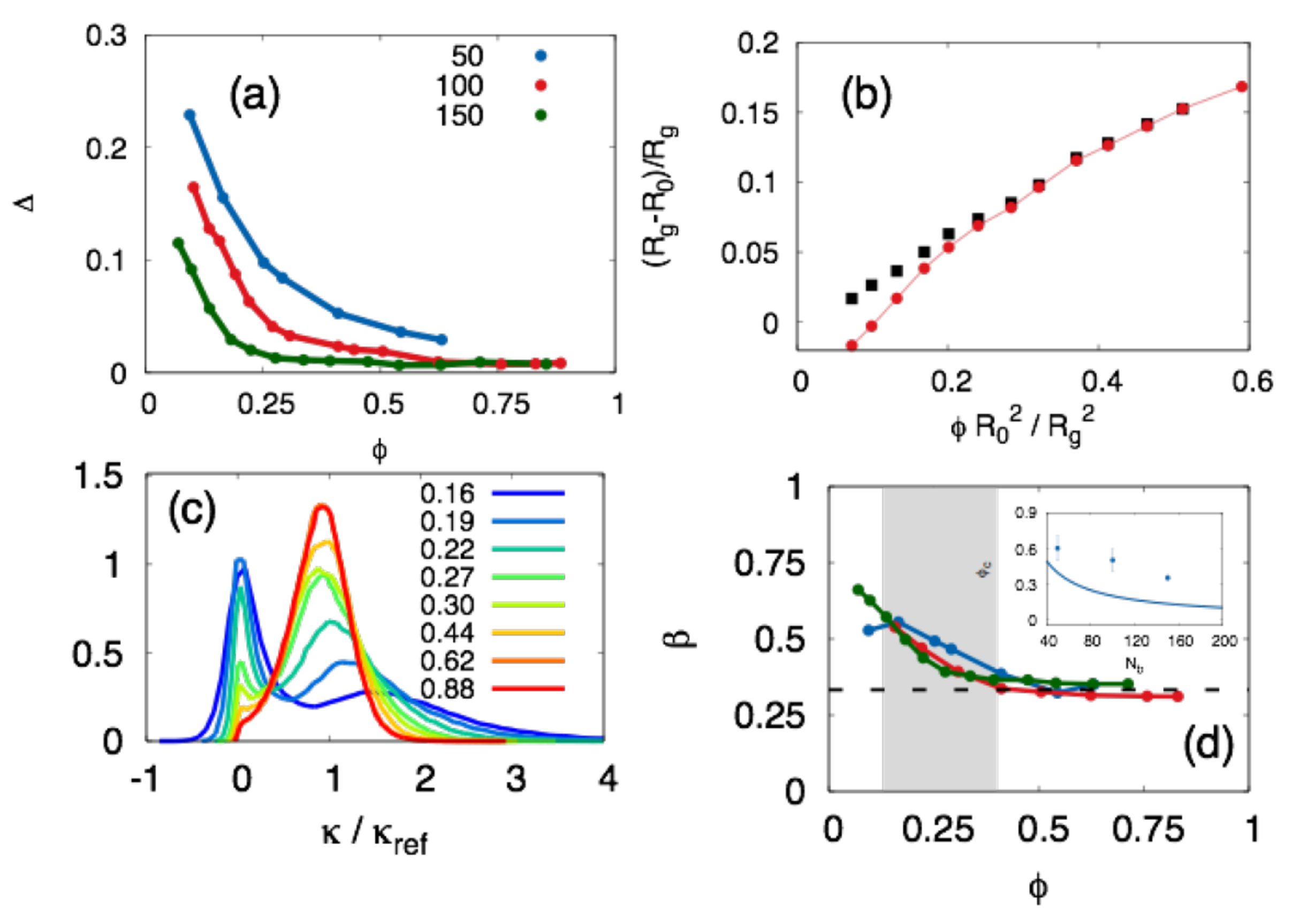}
\caption{ {\bf Membrane shape for elongated swimmers.} 
(a) Asphericity parameter for $N_b=50$ (blue symbols), $N_b=100$ (red symbols), and $N_b=150$ (green symbols). The lines are a guide to the eye. With increasing $\phi$ the vesicle approaches an average circular shape ($\Delta \sim 0$).
(b) The red circles are $(R_g-R_0)/R_g$, the red line is a guide to the eye, the black squares represent $P N_b / 2 \pi k$. The data are plotted as a function of the area fraction computed with respect the circle of radius $R_g$ for $N_b=150$.  (c) The probability distribution of the local curvatures undergoes a crossover from single to double peacked shape by increasing $\phi$ (in figure from $0.16$ (blue) to $0.83$ (red)). (d) To quantify the bimodal character of the distribution we look at the Sarle's bimodality coefficient $\beta$ as a function of $\phi$ for $N_b=50,100,150$ (blue, red and green), the black dashed line is $\beta$ for a Gaussian distribution. The grey area represents the estimated $\phi_c$ range. Inset: $\phi_c$ obtained from the decoupling between pressure and deformation (blue symbols) the line is the estimate of $\phi_c$ given by $\phi(N_b=N_s)$. 
}
\label{fig:fig3}       
\end{figure*}
\begin{figure*}[!t]
\centering
\includegraphics[width=1.\textwidth]{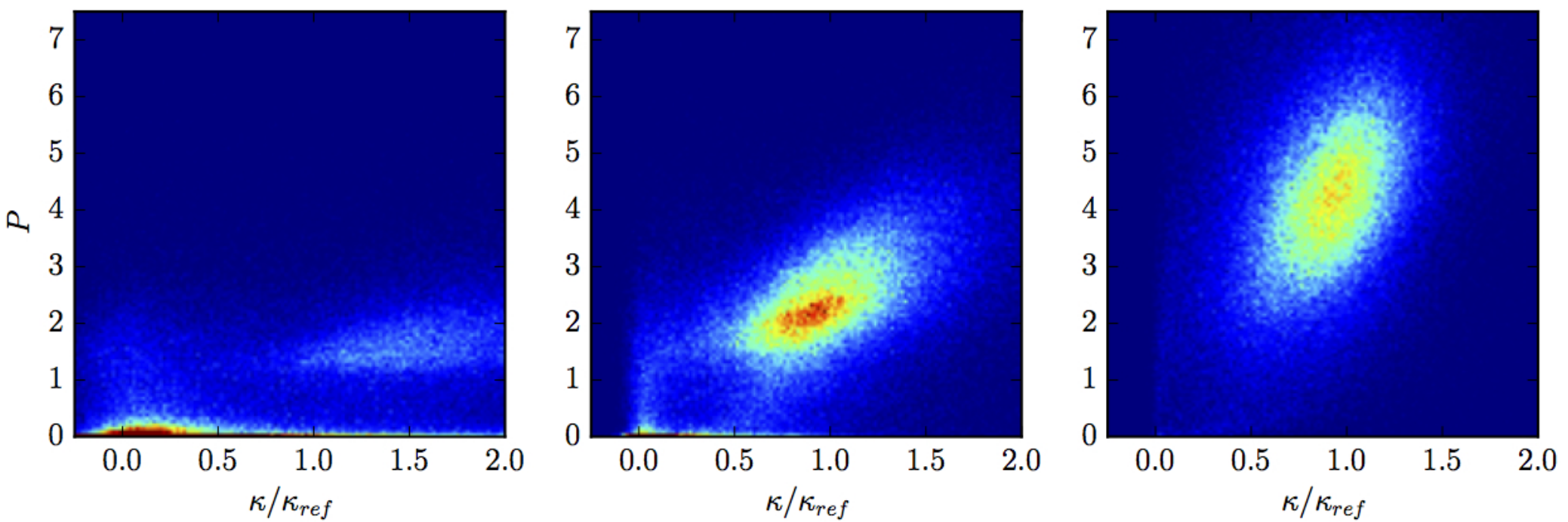}
\caption{{\bf Local curvature.} Joint probability distribution of the local curvature of the membrane $\kappa/\kappa_{ref}$
($\kappa_{ref}$ is the reference curvature of the circular free membrane) and the local pressure $P$
exerted by the swimmers on the membrane. The three panels refer to three different swimmers density,
$\phi=0.16$ (left), $0.51$ (middle) and  $0.83$ (right). Data correspond to the case of elongated swimmers and 
a membrane of $N_b=100$ beads.
}
\label{fig:fig2b}       
\end{figure*}

Now we consider elongated swimmers ($p=2$, aspect ratio $\alpha=1/2$).
In order to evaluate the impact of aligning forces on the membrane shape, we perform numerical
simulations of elongated active particles at almost the same area fractions $\phi$ of the previous case.
Again, to quantify the deviations of the vesicle from circular shape we display in Fig. \ref{fig:fig3}-a the
asphericity $\Delta$. We find that $\Delta$ approaches zero with increasing $\phi$, indicating that
at high density of swimmers the active vesicle approaches an average circular shape (see also the snapshot reported in the right panel of Fig. \ref{fig:fig1}).
On the contrary for small $\phi$ we observe strong deviation from a circle,
as displayed by the four snapshots
shown in Fig. \ref{fig:fig1}, left panel.

The radius of gyration increases with $\phi$, also for elongated swimmers, as one can see in Fig. \ref{fig:fig3}-b.
We observe, however, strong deviations from Eq. (\ref{press}) at low densities where $\Delta \neq 0$. 
This is not surprising since the right hand side of Eq. (\ref{press}) only holds when for circular membranes.  At high area fractions the vesicle shape becomes more 
circular ($\Delta\sim0$) and Eq. (\ref{press}) applies.
We stress that for elongated swimmers, where an aligning torque exists on the boundary, there is not 
an ideal active gas equation of state like (\ref{swim}) \cite{Solon15}. 

\subsection*{Local shape properties}
A useful characterization of the membrane shape is obtained 
by analyzing the distribution of local curvatures, $\mathcal{P}(\kappa/\kappa_{ref})$, with $\kappa_{ref}=R_0^{-1}$  the curvature
of the reference circular configuration, shown in 
Fig. \ref{fig:fig2}-c for spherical swimmers and in Fig. \ref{fig:fig3}-c for elongated swimmers.
Low density configurations are generically characterized by pinched spots, where both particle density and curvature are high, separated by straight, low curvature regions that are free from active particles (see the snapshot reported in Fig. \ref{fig:fig1}, left panel). 

Let us start by considering $\mathcal{P}(\kappa/\kappa_{ref})$ for elongated swimmers, where the asymmetry at low $\phi$ given by $\Delta(\phi)$ is more pronounced than for spherical swimmers.
The distribution changes from bimodal to unimodal with increasing packing fraction $\phi$. 
The bimodal distribution obtained at low $\phi$ corresponds to elongated vesicles. The
two peaks correspond to low curvature regions (where $\kappa\to0$ and the density of active swimmers is very low) and
high curvature regions (where swimmers accumulate and $\kappa > \kappa_{ref}$), respectively.
At large $\phi$  the vesicles are spherical on average ($\Delta\sim0$)
and the distribution of curvature exhibits a single peak. The finite width of the distribution 
measures the size of fluctuations about the mean shape 
with $\lambda_1 \sim \lambda_2$.

The bimodal character of the distribution can be quantified using the Sarle's bimodality coefficient $\beta=(\gamma^2+1)/k$,
where $\gamma$ is the skewness and $k$ the kurtosis of the distribution.
Fig. \ref{fig:fig3}-c shows the parameter $\beta$ reported as a function of swimmer density and for
three vesicle sizes. Deviations from the $1/3$ value, corresponding to a normal distribution,
observed at low swimmer density signals the appearance of the bimodality and associated elongated vesicle shape.
Particles tend to accumulate in small regions, enhancing the local membrane curvature, and 
leaving large parts of the membrane empty. The empty regions are flat and give a peak at a vanishing value of the local curvature. 
This results from a positive feedback mechanism by which a local fluctuation of particles density produces a local pressure increase that increases the local curvature on the flexible membrane. Since active particles tend to accumulate at convex boundaries, this local curvature increase drives  further accumulation of swimmers.

Fig. \ref{fig:fig2b} shows the joint probability density $\mathcal{P}(\kappa/\kappa_{ref},P)$. 
In the low density regime (left panel of Fig. \ref{fig:fig2b}),
flat regions of the membrane -- peak close to (0,0) in the figure -- coexist with 
highly curved regions -- lighter regions close to (1.5,1.5) in the figure 
(see also the snapshots reported in Fig. \ref{fig:fig1}).  
 By increasing the number of swimmers inside the vesicle the spot close to the origin 
disappears and a single broad peak at high $\kappa/\kappa_{ref}$ survives corresponding to uniform
curvature of the membrane -- see the snapshots of Fig. \ref{fig:fig1}, right panel, corresponding to the high particles density.

Similar results are obtained also for spherical swimmers, where the curvature distribution evolves from  double to single peaked with increasing area fraction $\phi$. In this case, however, this transition is sharper and occurs at lower values of $\phi$, and vesicles display a nearly circular shape in a wider range of area fractions.

The crossover from single to double peaked distribution of the membrane curvature
relies on the imbalance of swimmers along the boundaries. 
A rough estimate of the packing fraction $\phi_c$ at which the crossover takes place
can be obtained by the following argument. 
A membrane composed by $N_b$ beads of diameter $a$ has a length $a N_b$.
The minimum number of swimmers of thickness $a$ and length $\ell$  needed to 
uniformly cover the entire length of the membrane is 
$N_b$ (we suppose that the swimmers are pushing the membrane and that they are perpendicular to it).
The area fraction of swimmers is defined as $\phi=N_s a_{swim}/A_{ref}$,
where $A_{ref}= (N_b a_{swim})^2 / 4\pi$ is the area of the reference circular configuration of the free membrane.
The critical area fraction of swimmers is then  
$\phi_c=\phi(N_{s}=N_b)=p \pi^2/N_b$.
This  corresponds to the minimal swimmers density needed to obtain a uniform
distribution of pushing active particles along the membrane.
We obtain values of $\phi_c$ ranging $0.4$ to $0.13$ in the case of elongated swimmers,
and values from $0.2$ to $0.07$ in the case of spherical swimmers,
in agreement with the crossover regions observed in the behavior of $\beta$ (Fig. \ref{fig:fig2}-d and Fig. \ref{fig:fig3}-d
where the grey area represents the $\phi_c$ range).

A numerical estimate of $\phi_c$ is obtained as follows. 
When $\Delta \neq 0$, Eq. (\ref{press}) does not hold and the relative displacement $(R_g-R_0)/R_0$
is not proportional to the average pressure exerted by the active particles. We define $\phi_c$ as
the value of $\phi$ where Eq. (\ref{press}) begins to hold. 
In the inset of Fig. \ref{fig:fig2}-d the
line is the estimate of $\phi_c$ given by $\phi(N_b=N_s)$ and the symbols are the numerical values (spherical swimmers)
 obtained looking at the deviation from Eq. (\ref{press}). 
The curve reproduces quite well the data.
Different is the situation for the elongated swimmers (inset in Fig. \ref{fig:fig3}-d), where
the numerical estimate lies above the curve $\phi(N_b=N_s)$, i. e., the steric effect is not enough to justify the rise in $\phi_c$.

\subsection*{ Cell migration}

Flexible vesicles do not just fluctuate in shape but, at the same
time, perform a random walk under the action of the fluctuating
force arising from the combined action of swimmers' propelling
forces.  
The case of spherical swimmmers is particularly remarkable
since it can be worked out analytically.  Since swimmers
and passive beads have the same size and mobility, the center of mass
velocity $\mathbf V_{cm}$ is given by
\BEQ\label{vcdm}
{\mathbf V}_{cm}=\frac{1}{N_b+N_s}\left[\sum_n^{N_b}
  {\mathbf V}_n+\sum_j^{N_s} {\mathbf v}_j\right]=\frac{\mu}{N_b+N_s} \sum_j\mathbf f_j \, ,
\EEQ
where $\mathbf V_n$ and $\mathbf v_j$ are respectively the velocities of a membrane bead and a swimmer. 
The sum of all interaction forces has to vanish so that only the sum over propelling forces 
$\mathbf f_j$  survives in the last term.
Therefore the center of mass moves as a body of reduced mobility
$\mu/(N_b+N_s)$ under the action of the total propelling force on the
swimmers.  The corresponding velocity-velocity correlation function is then given by
\begin{equation}
\langle\dot{\mathbf R}_{cm}(0)\cdot\dot{\mathbf R}_{cm}(t)\rangle=
\frac{\mu^2}{\left(N_s+N_b\right)^2}\sum_{i,j}^{N_s}\langle\mathbf
f_i(0)\cdot\mathbf f_j(t)\rangle\label{cdm} \, .
\end{equation}
For spherical swimmers, propelling forces only reorient due to tumbles
and are therefore uncorrelated so that
\begin{equation}
\sum_{i,j}^{N_s}\langle\mathbf f_i(0)\cdot\mathbf f_j(t)\rangle= N_s
\langle\mathbf f(0)\cdot\mathbf f(t)\rangle=\frac{N_s v^2}{\mu^2}
e^{-\lambda |t|}
\end{equation}
The mean square displacement (MSD) is obtained by a double time
integration of (\ref{cdm}), with the result
\bea
\label{msd} 
\nonumber \langle|\Delta\mathbf R_{cm}(t)|^2\rangle&=& \frac{N_s
  v^2}{\left(N_s+N_b\right)^2}\int_0^t dt^\prime\int_0^t
dt^{\prime\prime}e^{-\lambda |t^\prime-t^{\prime\prime}|}\\ 
&=&\frac{4 D_v}{\lambda}\left( \lambda t - 1 + e^{-\lambda t}\right)
\eea
\begin{figure*}[!t]
\centering
\includegraphics[width=.75\textwidth]{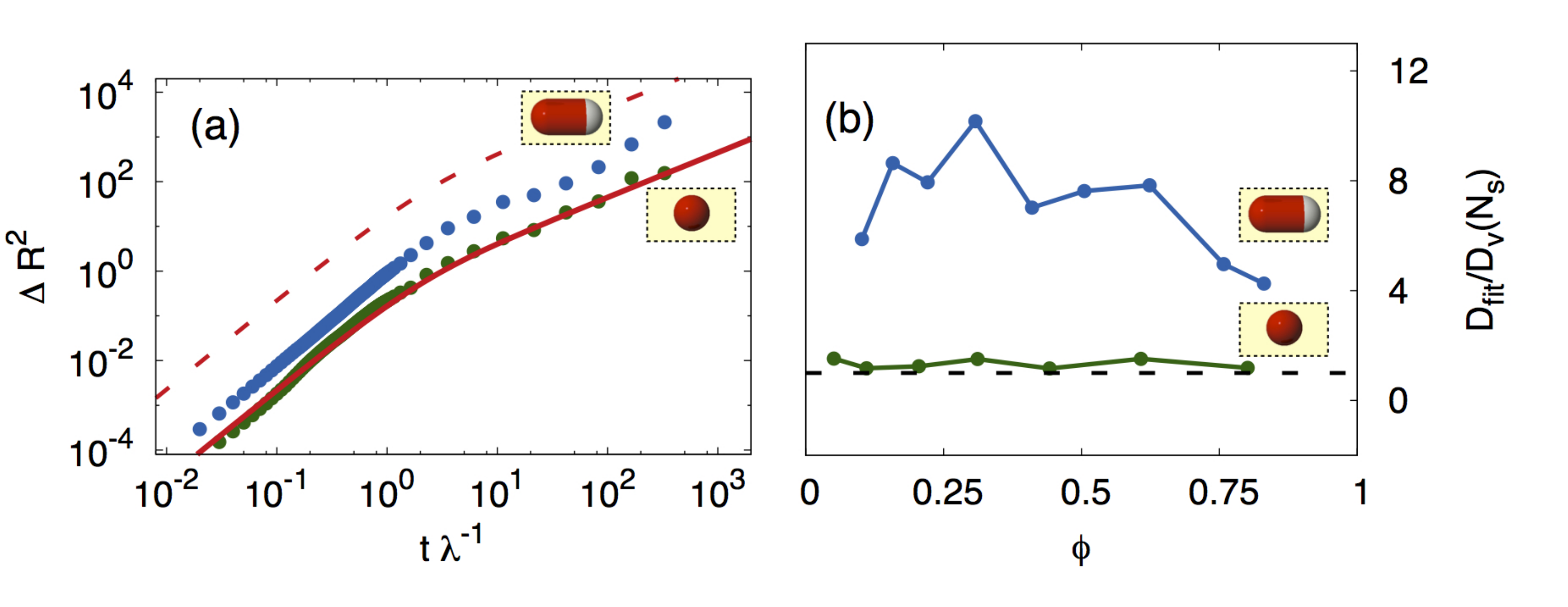}
\caption{ {\bf Vesicle motion.} (a) Mean square displacement of the vesicle center of mass for $\phi=0.16$ and $N_b=100$.
Data correspond to the cases of elongated swimmers (blue symbols) and spherical swimmers (green symbols). 
The red curve is the theoretical prediction given by eq. (\ref{msd}). The red dashed curve is the theoretical mean square displacement of a free run-and-tumble particle. (b) Diffusion coefficient normalized to the reduced value $D_v(N_s)$ for elongated (blue symbols) and spherical (green symbols) swimmers as a function of swimmer density $\phi$ ($N_b=100$). The parameters are obtained by fitting the data with eq. (\ref{msd}). The black dashed line is $D_{fit}/D_v=1$.}
\label{fig:fig5}       
\end{figure*}
The MSD of the center of mass of the vesicle is then given by the
MSD of an individual swimmer, with the single swimmer diffusivity $D=v^2/2 \lambda$ 
replaced by the reduced value $D_v=D N_s/\left(N_s+N_b\right)^2$.  The MSD
of a free swimmer \cite{Angelani13} and of a vesicle filled with spherical swimmers
are shown in Fig. \ref{fig:fig5}-a together with the formula
(\ref{msd}). 
The case of non spherical swimmers is more complex due
to the rotational couplings between propelling forces induced by
anisotropic interactions. Still the calculated MSD can be fitted with
formula (\ref{msd}) leaving both $D_v$ and $\lambda$ as free fitting
parameters. In this case, however, we expect that due to anisotropic
interactions, correlations between $\mathbf f_j$ will arise whose
relaxation is not solely driven by the tumbling rate $\lambda$ but can
occur on longer time scales. The obtained fitting parameters confirm
those expectation giving $\lambda_{fit}\sim0.3\lambda$.

The fitted diffusion coefficients as a function of particles density are reported in 
Fig. (\ref{fig:fig5}-b) for both spherical and elongated swimmers. 
As expected, the diffusion coefficient in the spherical case is given by the reduced value $D_v$.
In the case of elongated swimmers the vesicle diffusivity is much larger due to a longer persistence of propelling forces arising from locally aligned configurations of swimmers.

\section*{Discussion}

Understanding the properties of active matter in confined geometries is of great importance
not only for basic science, but also for possible practical applications,
for example in micro bio-mechanics, where synthetic autonomous  self-propelled objects could be 
used as drug-delivery agent or for mechanical actuation.
Previous studies have focused on the behavior of active particles in the presence of
rigid obstacles or confined by stiff boundaries \cite{Smallenburg15,Mallory14,Fily15,Fily14}. 
In this paper we explore the  shape changes and spontaneous migration of a flexible vesicle filled with active particles.   
We find strong fluctuations of the vesicle's shape, changing from circular to elongated with decreasing 
number of enclosed particles. 
The transition between these two regimes is associated with the crossover of the distribution of the local 
curvatures $\mathcal{P}(\kappa)$
 from single-peaked to bimodal.
The  observed shape deformation is driven by
the accumulations of active particles in the high
curvature regions, which  has been observed also in the case of non interacting
Active Brownian particles under strong confinement \cite{Fily14}. Elongated swimmers enhance shape deformations because alignment tends to increase particle accumulation in high curvature regions. 

We have recently become aware of a study similar to ours investigating shape fluctuations in 2D flexible vesicles 
filled with  \textit{spherical} Active Brownian particles \cite{Tian15}. 
Although in this work particles' trajectories are randomized by rotational diffusion while we use run-and-tumble dynamics,  both our work and Ref.  \cite{Tian15} find similar robust shape fluctuations induced by the active particles. The transition from elongated to circular vesicle shape that we observed by increasing density of enclosed swimmers is found in Ref.  \cite{Tian15} upon decreasing the particles' propelling force. 

We also show that the filled vesicle effectively behaves like an active object, with
 exponentially correlated random motion, whose properties
are strongly dependent on the shape and density of the self-propelled pushing particles inside. 
In the case of spherical swimmers we can calculate the diffusion coefficient $D_v$ and the 
correlation time $\tau$ of the persistent random walk of the filled vesicle, that can be described in terms of an effective temperature that depends on the number of enclosed swimmers. The migratory properties of 
the cell are determined entirely by the motility of the active particles.

We additionally examine the behavior of vesicles filled with elongated particles that was not considered in Ref.  \cite{Tian15}.
In this case the diffusion coefficient of the whole vesicle is about one order of magnitude greater than that of the
spherical case and it is a non-monotonic function of the swimmers density,
reaching a maximum value near the critical packing fraction $\phi_c$ controlling the crossover from single to double peaked distribution 
of the membrane curvature.

The  behavior of vesicles filled with active particles bear some resemblance with the directed migration of Eukaryotic cells, as observed for instance in wound healing assays or in the presence of chemotactic cues. In these situations cells become polarized and perform directed random walks advancing preferentially toward or away from chemical stimuli \cite{Swaney10}  or towards regions void of other cells~\cite{Kim13}. 
Our work shows (see Fig. \ref{fig:fig5}b)  that vesicle migration is most effective when driven by elongated particles that indeed induce a net polarization of the vesicle, as observed in the chemotactic motion of living cells.  It would be interesting to study the effect
of chemotaxis on our model by considering a
space-varying tumbling rate $\lambda(\rr)$ which depends on an external chemotactic field $c(\rr)$.



\section*{Acknowledgements}
RDL acknowledges funding from the European Research Council under the European Union's
Seventh Framework Programme (FP7/2007-2013)/ERC grant agreement No. 307940.
 MP and MCM were supported by the Simons Foundation  Targeted Grant in the Mathematical Modeling of Living Systems Number: 342354 and by the Syracuse
Soft Matter Program.
 MCM also acknowledges  support
by the National Science Foundation through award DMR-1305184. 

\bibliography{biblio}

%


\end{document}